\newcommand{\vecu}{\mbox{\boldmath$u$}}

\newcommand{\vecE}{\mbox{\boldmath$E$}}
\newcommand{\vecB}{\mbox{\boldmath$B$}}
\newcommand{\vecS}{\mbox{\boldmath$S$}}
\newcommand{\vece}{\mbox{\boldmath$e$}}
\newcommand{\vecf}{\mbox{\boldmath$f$}}
\newcommand{\vecnl}{\mbox{\boldmath$0$}}
\newcommand{\dfd}{{\rm d}}
\newcommand{\vecr}{\mbox{\boldmath$r$}}
\newcommand{\vecv}{\mbox{\boldmath$v$}}

\newcommand{\matu}{\mbox{\textsf{\textbf u}}}
\newcommand{\matF}{\mbox{\textsf{\textbf F}}}
\newcommand{\matH}{\mbox{\textsf{\textbf H}}}
\newcommand{\matone}{\mbox{\textsf{\textbf 1}}}

\documentclass[12pt,a4paper]{article}
\usepackage{graphicx}

\title{The pushing force of a propagating electromagnetic wave}
\author{Hanno Ess\'en\\
Department of Mechanics \\Royal Institute
of Technology  \\ S-100 44 Stockholm, Sweden}
\begin{document}
\maketitle
\begin{abstract}
The effect of the electrodynamic forces on a charged particle in a
propagating plane electromagnetic wave is investigated. First it
is pointed out that for constant fields fulfilling the radiation
condition there will be an acceleration in the direction of the
Poynting vector. When oscillating fields are considered the
Lorentz force on the particle only causes a drift, with constant
average velocity, in the direction of propagation of the wave,
i.e.\ the direction of the Poynting vector. Finally, when the
radiative reaction (radiation damping) force is added the result
is again an acceleration in the direction of wave propagation.
\vskip 1cm \noindent PACS classification numbers: 03.50.De,
41.60.-m, 41.75.Jv
\end{abstract}

\newpage
\section{Introduction}
Recently Professor Thomas Gold \cite{gold} has published a
manuscript on the Web stating that radiation pressure from the Sun
does not exist, or at least that it cannot be used for propulsion
with solar sails, as has been suggested \cite{friedman}. I do not
understand Gold's, mainly thermodynamic, arguments but the
phenomenon of radiation pressure is well established theoretically
and experimentally so the manuscript is somewhat surprising.
Usually the force of radiation pressure is explained by discussing
how photons carry momentum which is either absorbed or reflected
when impinging on a body. Here we will point out and elucidate the
less known fact that the electrodynamic forces (Lorentz force and
radiation damping force) on a charged particle from a plane
electromagnetic wave accelerates the particle in the direction of
propagation of the wave. This, at least, demonstrates that
radiation pressure is an immediate consequence of the relativistic
equation of motion of a charged particle in an electromagnetic
field.

For simplicity of notation we will use Gaussian units. The
relativistic equation of motion for a charged particle of charge
$q$ and mass $m$, in an external electromagnetic field
$F_{ab}=(\vecE, \vecB)$ (with the notation of Landau and Lifshitz
\cite{landau}) is given by,
\begin{equation}
\label{eq.lorentz.force.tens} m\frac{\dfd u^a}{\dfd \tau}
=\frac{q}{c} F^a_{\;\; b} u^b ,
\end{equation}
if we, for the time being, neglect radiative reaction. Here $ u^a
=(1/c) \dfd x^a/\dfd \tau =\gamma (1, \vecv/c)=(u^0,\vecu)$, where
$x^a=(ct,\vecr)$, and $\tau$ is proper time, $\dfd\tau=\gamma \dfd
t$, and, finally, $\gamma=u^0 = 1/\sqrt{1-v^2/c^2}$. Later we will
add $g^a$, the radiation damping force, on the right hand side.
This is a correction needed since an accelerated charge radiates
and this results in a reaction force. If one introduces the new
independent variable
\begin{equation}
\zeta =\frac{q\tau}{mc},
\end{equation}
one can write Eq.\ (\ref{eq.lorentz.force.tens}) in the form
\begin{equation}
\label{eq.lorentz.force.three} \left( \frac{\dfd \gamma}{\dfd
\zeta}, \frac{\dfd \vecu}{\dfd \zeta} \right)
 =\left( \vecu\cdot\vecE,\,  \gamma \vecE + \vecu \times \vecB \right)
\end{equation}
for the time and space components respectively.

\section{Motion due to the Lorentz force of constant fields}
The general solution of Eq.\ (\ref{eq.lorentz.force.three}) for
constant electromagnetic fields $F_{ab}=(\vecE, \vecB)$ is known
and has been discussed extensively in the literature. In
particular one can recommend the studies by Salingaros
\cite{salingaros1,salingaros2}. Other illuminating contributions
are by  Hyman \cite{hyman} and by Mu\~nos \cite{munos}.

First we note that the case of radiation is very special and
differs from the general case. For constant $\vecE$ and $\vecB$
one can in general make a Lorentz transformation to a reference
frame in which these two vectors are parallel. Then the Poynting
vector,
\begin{equation}
\label{eq.poynting.vec} \vecS = \frac{c}{4\pi}\vecE \times \vecB ,
\end{equation}
which represents the flux density of momentum in the
electromagnetic field, becomes zero. In that frame therefore there
can be no radiation pressure force. In such general fields charged
particles will have a drift velocity equal to the velocity of the
frame in which the fields are parallel. Radiation fields on the
other hand are characterized by
\begin{eqnarray}
\vecE\cdot\vecB=0 ,\\
|\vecE|=|\vecB|,
\end{eqnarray}
and for such fields there is no such reference frame of zero
Poynting vector. The general solution of Eq.\
(\ref{eq.lorentz.force.three}) for constant $\vecE$ and $\vecB$
fulfilling the radiation conditions, for a particle starting at
rest, is given by Salingaros \cite{salingaros2} and is
\begin{equation}
\label{salingaros.sol} \label{eq.const.rad.case.sol} \vecu =\zeta
\vecE +\frac{1}{2}\zeta^2 \vecE \times \vecB ,\;\;\; \gamma=1
+\frac{1}{2}\zeta^2 |\vecE|^2 .
\end{equation}
The meaning of this result is that, for small $\zeta$ the velocity
will be essentially parallel (or anti-parallel) to the electric
field as intuition demands, but for large $\zeta$ the velocity
will become more and more parallel (never anti-parallel) to the
Poynting vector $\vecS =(c/4\pi) \vecE \times \vecB$.

\section{Motion due to the Lorentz force of of propagating wave}
The result (\ref{eq.const.rad.case.sol}) by Salingaros and its
interpretation does not survive when the field is allowed to have
an oscillatory time dependence, as we will now show. We start by
rewriting Eq.\ (\ref{eq.lorentz.force.tens}),
\begin{equation}
\label{eq.lorentz.force.tens.zeta} \frac{\dfd u^a}{\dfd \zeta} =
F^a_{\;\; b} u^b ,
\end{equation}
on matrix form, as follows
\begin{equation}
\label{eq.lorentz.force.matrix} \frac{\dfd \matu}{\dfd\zeta} =
\matF(\zeta) \matu ,
\end{equation}
and we will allow $\matF$ to depend on time via $\zeta$. Here the
components of the matrices $\matF$ and $\matu$ are given by
\begin{equation}
\label{eq.matF.fields} \matF= \left( \begin{array}{cccc}
0 & E_x & E_y & E_z \\
E_x & 0 & B_z & - B_y \\
E_y & -B_z & 0 & B_x \\
E_z & B_y & -B_x & 0
\end{array} \right), \;\; \mbox{and} \; \;
\matu= \left( \begin{array}{c} u^0 \\ u_x \\u_y \\ u_z \end{array}
\right),
\end{equation}
respectively. We now specialize to the radiation case and chose
the x-axis in the direction of $\vecE$ so that $\vecE= E(\zeta)
\vece_x$. If also chose the z-axis as the direction of the
Poynting vector we must have $\vecB$ along the y-axis and of the
same length as $\vecE$, so that $\vecB= E(\zeta) \vece_y$. The
Poynting vector is then $\vecS =\frac{c}{4\pi}  E^2(\zeta)
\vece_z$, and the $\matF$-matrix is given by
\begin{equation}
\label{eq.matF.fields.rad} \matF= \left( \begin{array}{cccc}
0 & E(\zeta) & 0 & 0 \\
E(\zeta) & 0 & 0 & - E(\zeta) \\
0 & 0 & 0 & 0 \\
0 & E(\zeta) & 0 & 0
\end{array} \right) \equiv E(\zeta)\matH ,
\end{equation}
where we have defined the matrix
\begin{equation}
\label{eq.matH} \matH= \left( \begin{array}{cccc}
0 & 1 & 0 & 0 \\
1 & 0 & 0 & - 1 \\
0 & 0 & 0 & 0 \\
0 & 1 & 0 & 0
\end{array} \right) .
\end{equation}
We have now found that, in the radiation case, the equation of
motion, Eq.\ (\ref{eq.lorentz.force.matrix}), becomes
\begin{equation}
\label{eq.lorentz.force.matrix.rad} \frac{\dfd \matu}{\dfd\zeta} =
\left[ E(\zeta)\matH \right] \matu .
\end{equation}
Thus, differentiating $\vecu$ with respect to $\zeta$ multiplies
$\vecu$ with a matrix. The general solution of this equation is
given by
\begin{equation}
\label{eq.lorentz.force.matrix.rad.sol} \matu(\zeta) = \exp\left[
\int_0^{\zeta} E(\eta) \dfd\eta\, \matH \right] \matu_0 .
\end{equation}
To get an explicit solution we put
\begin{equation}
\label{eq.def.f} f(\zeta) =  \int_0^{\zeta} E(\eta) \dfd\eta ,
\end{equation}
and use the series expansion of the exponential. For this we need
the powers of the matrix $\matH$ and these are: $\matH^0=\matone$,
the four by four unit matrix, $\matH^1 = \matH$,
\begin{equation}
\label{eq.matH2} \matH^2= \left( \begin{array}{cccc}
1 & 0 & 0 & -1 \\
0 & 0 & 0 & 0 \\
0 & 0 & 0 & 0 \\
1 & 0 & 0 & -1
\end{array} \right) .
\end{equation}
All higher powers of $\matH$ are zero matrices. We thus find that
\begin{equation}
\label{eq.lorentz.force.matrix.rad.sol1} \matu(\zeta) =
\left[\matone + f(\zeta)\matH + \frac{1}{2} f^2(\zeta)
 \matH^2 \right] \matu_0 .
\end{equation}
We now assume that the initial condition is $\tilde{\matu}_0=(1\;
0\; 0\; 0)$, i.e.\ that the particle is at rest. Explicit
calculation then gives
\begin{equation}
\label{eq.lorentz.force.matrix.rad.sol.expl} \matu(\zeta) = \left(
\begin{array}{c} 1+\frac{1}{2}f^2(\zeta) \\ f(\zeta)\\0 \\
\frac{1}{2}f^2(\zeta) \end{array} \right).
\end{equation}
If we take $E(\zeta)=$ const.\ we recover Salingaros' solution
(\ref{salingaros.sol}).

If we instead assume that we have a simple harmonic wave, so that
$E(\zeta) = E_1 \cos(w\zeta)$ and $f(\zeta)=E_1\sin(w\zeta)/w$. We
then find that the time average of the four velocity
(\ref{eq.lorentz.force.matrix.rad.sol.expl}) becomes
\begin{equation}
\label{eq.time.aver} < \matu(\zeta)>_{\zeta} \equiv \frac{w}{2\pi}
\int_0^{2\pi/w} \matu(\zeta) \dfd\zeta =\left(
\begin{array}{c} 1+\frac{E_1^2}{4w^2} \\ 0 \\0 \\
\frac{E_1^2}{4w^2} \end{array} \right).
\end{equation}
The acceleration in the direction of the Poynting vector (the
z-axis) that we found in the case of constant fields has now
become a drift with constant average speed (originally calculated
by McMillan \cite{mcmillan}). The speed of this drift is larger
the smaller the frequency $w$ is, but there is no acceleration and
thus no average force in the direction of the Poynting vector. For
a recent discussion of this problem, see McDonald and Shmakov
\cite{mcdonald}.

\section{The radiative reaction force from propagating plane wave}
So far we have neglected the radiative reaction force and used the
equation of motion, Eq.\ (\ref{eq.lorentz.force.tens}). To make it
more accurate we must add the four force $g^a$ on the right hand
side. Now we investigate the form of $g^a$ and how it will modify
the solutions found in the previous sections.

The origin of this force is the electromagnetic radiation that an
accelerated charged particle sends out. This radiation carries
energy and momentum and there is therefor a reaction force on the
particle itself. One can show that the force, due to dipole
radiation, should be $\vecf=\frac{2q^2}{3c^3}\ddot {\vecv}$. The
four vector form of this should give the four force $g_0
^a=\frac{2q^2}{3c^3} \frac{\dfd^2 u^a}{\dfd \tau^2}$. A four force
must for kinematic reasons fulfill $g^a u_a =0$, the four scalar
product with the four velocity must be zero. This is achieved by
modifying $g_0^a$ by subtracting the component along the four
velocity: $g_1^a = g_0^a - (g_0^b u_b) u^a$. Since $u_a u^a=1$
this gives the desired property. Finally one can insert the
expression for $\frac{\dfd^2 u^a}{\dfd \tau^2}$ obtained by
differentiating Eq.\ (\ref{eq.lorentz.force.tens}), and also the
expression for $\frac{\dfd u^a}{\dfd \tau}$, from the same
equation. The force $g^a$ should, after all, be a perturbation
compared to the Lorentz force. Thus one obtains
\begin{eqnarray}
 g^a =\frac{2q^2}{3c^3} \left\{ \frac{q}{mc}
\left[ \frac{\dfd F^a_{\;\; b}}{\dfd \tau} u^b + F^a_{\;\; b}
\left( \frac{q}{mc}  F^b_{\;\; c} u^c \right) \right] \right\}
\\-\frac{2q^2}{3c^3} \left\{ \frac{q}{mc} \left[ \frac{\dfd
F^b_{\;\; c}}{\dfd \tau} u^c + F^b_{\;\; c} \left( \frac{q}{mc}
F^c_{\;\; d} u^d \right) \right] \right\}u_b u^a .
\end{eqnarray}
Some algebra, and use of the fact that $F^{ab}u_a u_b=0$ (due to
the anti-symmetry of $F^{ab}$), turns this into
\begin{equation} \label{eq.rad.damp.g} g^a =\frac{2q^3}{3mc^4}
 \frac{\dfd F^a_{\;\; b}}{\dfd \tau}u^b  +
 \frac{2q^4}{3m^2c^5} \left(  F^a_{\;\; c}
  F^c_{\;\; b}    - F^d_{\;\; c}
 F^c_{\;\; b}  u_d u^a  \right)u^b.
\end{equation}
The corresponding three dimensional force $\vecf$ can be obtained
by noting that, with our conventions, $g^a = (\gamma/c)
(\vecf\cdot\vecv/c, \vecf)$.

Landau and Lifshitz \cite{landau} (\S 76, Problem 2) give the
three dimensional form for this force as follows
\begin{eqnarray}
\label{eq.three.dim.rad.react.LL2.o} \vecf = \vecf_o + \vecf_s
=\frac{2}{3}\frac{q^3}{mc^3} \gamma \left( \frac{\dfd \vecE }{\dfd
t} + \frac{\vecv}{c}\times \frac{\dfd \vecB }{\dfd t} \right) \\
\nonumber
 + \frac{2}{3}\frac{q^4}{m^2
c^4} \left\{  \left[ \vecE\times\vecB + \left(
\frac{\vecv}{c}\times \vecB
 \right) \times \vecB + \left( \frac{\vecv}{c}\cdot \vecE
 \right) \vecE \right] \right. \\
 \label{eq.three.dim.rad.react.LL2.s} \\
 \nonumber
 \left. -\frac{\vecv}{c} \gamma^2  \left[ \left( \vecE +\frac{\vecv}{c}
\times \vecB \right)^2 - \left( \frac{\vecv}{c}\cdot \vecE
 \right)^2 \right] \right\}.
\end{eqnarray}
Here $\frac{\dfd}{\dfd t} = \frac{\partial}{\partial t} +
\vecv\cdot\nabla$ and $\gamma=1/\sqrt{1-v^2/c^2}$. We will now
analyze the effect of this force for a charge in an oscillatory
plane wave.

The first term, $\vecf_o$, in line
(\ref{eq.three.dim.rad.react.LL2.o}) is then of an oscillatory
character. It can, in fact, be combined with the Lorentz force,
\begin{eqnarray}
\label{eq.lor.and.o.comb} \vecf_l(t) + \vecf_o(t) =\left(
q\vecE(t) +q \frac{\vecv}{c}\times\vecB(t)\right) +
\frac{2}{3}\frac{q^3}{mc^3} \gamma \left( \frac{\dfd \vecE }{\dfd
t} + \frac{\vecv}{c}\times
\frac{\dfd \vecB }{\dfd t} \right)\\
\approx q\vecE(t+\epsilon_t) +q
\frac{\vecv}{c}\times\vecB(t+\epsilon_t)= \vecf_l(t+\epsilon_t),
\end{eqnarray}
and we see that the effect of this term is to make a tiny time
translation of the Lorentz force with
$\epsilon_t=\frac{2}{3}\frac{q^2}{mc^3}\gamma$. This essentially
is the time needed for light to pass across the classical electron
radius. This is unlikely to have any noticeable physical effects.

The essential radiation damping force is thus $\vecf_s$. All its
terms (\ref{eq.three.dim.rad.react.LL2.s}) are quadratic in the
fields so there is hope that this force, $\vecf_s$, produces some
net work on the particle. In the rest frame ($\vecv=\vecnl$) we
simply get,
\begin{equation}
\label{eq.three.dim.rad.react.v-zero}  \vecf_s (\vecv=\vecnl) =
\frac{2}{3}\frac{q^4}{m^2 c^4} \vecE\times\vecB,
\end{equation}
and therefore the particle starts to accelerate in the direction
of the Poynting vector. What happens when it picks up speed?

Assume that the field is of the type in Eq.\
(\ref{eq.matF.fields.rad}) with oscillatory $E=E(\zeta)$. Since
$\vecE= E \vece_x$, $\vecB= E\vece_y$, and $\vecE\times\vecB = E^2
\vece_z$, some algebra then shows that our radiation reaction
force, quadratic in the fields, is
\begin{eqnarray}
\label{eq.fs.for.rad.field} \vecf_s = \frac{2}{3}\frac{q^4}{m^2
c^4}E^2 \left(1-\frac{v_z}{c} \right) \left[\vece_z
-\gamma^2\left(1-\frac{v_z}{c}\right) \frac{\vecv}{c}
 \right].
\end{eqnarray}
We see that the effect of this force is always to accelerate the
particle in the direction, $\vece_z$, of the Ponting vector. The
dissipative, $-\vecv$, term is one order higher in $v/c$ and will
thus only become important near relativistic speeds. The force
goes to zero when $\vecv \rightarrow c\vece_z$, but it does not
change direction.

Surprisingly we find that this force, which usually is called the
radiation damping force, does not damp anything in this case.
Instead its effect is the acceleration of a charged particle in
the direction of propagation of a travelling electromagnetic wave.

\section{Conclusions}
Above we have presented an elegant matrix method that yields an
exact solution for the motion of a particle in an electromagnetic
radiation field with a given time dependence. This generalizes
some older results for constant fields of Salingaros
\cite{salingaros1,salingaros2}. The concise explicit formula of
Eq.\ (\ref{eq.fs.for.rad.field}) for the radiation damping force
in a plane wave might be also be a new contribution. The main aim
of this article has been to illuminate the way in which classical
electrodynamics leads to a pushing of charges along the direction
of of wave propagation.

 Just as in solar sailing, this is a very small force, but
in space where there normally are no dissipative forces, even a
small force can have very large effect in the long run. Note that
there is no need for a force to be stronger than gravity to
accelerate things outwards from the sun (a common
misunderstanding). A typical object orbits the sun in a stationary
ellipse. When a small force is added this ellipse will gradually
change, and with time, these changes can become very large.

\end{document}